\documentclass[dvips,12pt]{article}
\usepackage{graphicx}
\begin{document}
\markboth{}{Papapetrou Equation for Constrained Electromagnetic
Field }

\title{Variational Principles for Constrained Electromagnetic
Field and Papapetrou Equation}
\date{}

\author{\footnotesize Muminov~A.T. \\
\footnotesize Ulug-Beg Astronomy Institute, Astronomicheskaya~33,
Tashkent 100052,
Uzbekistan\\\footnotesize{amuminov@astrin.uzsci.net}}

\maketitle

\begin{abstract}\footnotesize
In our previous article \cite{za} an approach to derive Papapetrou
equations for constrained electromagnetic field was demonstrated
by use of field variational principles. The aim of current work is
to present more universal technique of deduction of the equations
which could be applied to another types of non-scalar fields. It
is based on Noether theorem formulated in terms of Cartan'
formalism of orthonormal frames. Under infinitesimal coordinate
transformation the one leads to equation which includes volume
force of spin-gravitational interaction. Papapetrou equation for
vector of propagation  of the wave is derived on base of the
equation. Such manner of deduction allows to formulate more
accurately the constraints and clarify equations for the potential
and for spin.
\end{abstract}

\def \vc {\vec}
\newcommand{\T}{\tilde }
\newcommand{\lr}{\left(}
\newcommand{\rr}{\right)}
\newcommand{\lfc}{\left\{}
\newcommand{\rtc}{\right\}}
\newcommand{\lta}{\left\langle}
\newcommand{\rta}{\right\rangle}
\def \om {\omega}
\def \wd {\wedge}
\def \ex {\exp{i(\omega t-m\vphi)}}
\def \va {\vc A }
\def \vac {\bar{\vc A }}
\def \ca {\bar{A}}
\def \vcx {\dot{\vc x }}
\def \dx {\delta\vc x }
\def \vca {\dot{\vc A }}
\def \vcca {\dot{\bar{\vc A }}}
\def \ccn {\mbox{C.C. }}
\newcommand{\lcom}[1]{{\left[#1\right.}}
\newcommand{\rcom}[1]{{\left.#1\right]}}
\newcommand{\lacm}[1]{{\left\{#1\right.}}
\newcommand{\racm}[1]{{\left.#1\right\}}}
\newcommand{\cc}[1]{\bar{#1}}
\newcommand{\beql}[1]{\begin{equation}\label{#1}}
\newcommand{\eeq}{\end{equation}}
\newcommand{\half}{\frac{1}{2}}
\newcommand{\quart}{\frac{1}{4}}
\newcommand{\too}{\longrightarrow}
\newcommand{\eps}{\varepsilon}
\newcommand{\vphi}{\varphi}
\newcommand{\fracm}[2]
{{\displaystyle#1\over\displaystyle#2\vphantom{{#2}^2}}}
\newcommand{\prt}{\partial\,}
\def \oh {{\tiny 1/2}\,}
\def \moh {{-1/2}}
\def \as {\,^*}
\def \elm {electromagnetic }
\def \om {\omega}
\def \Om {\Omega}
\def \ex {\exp{i(\omega t-m\vphi)}}
\def \man {{\cal M}}
\def \dm {{differentiable manifold }}
\def \vn {{\bf n}}
\def \cdt {\hspace{0.1em}\cdot\hspace{0.1em}}
\def \csdt {\hspace{0.35em}\cdot\hspace{0.35em}}
\def \fdt {\hspace{0.4em}}
\def \tch {\;\grave{}}
\newcommand{\con}[2]{\omega_{#1\cdt}^{\fdt#2}}
\newcommand{\gcn}[2]{\,\grave{}\omega_{#1\cdt}^{\fdt#2}}
\newcommand{\cur}[2]{\Omega_{#1\cdt}^{\fdt#2}}
\newcommand{\tdu}[3]{#1_{\ldots #2\ldots}^{\mdots\ldots #3\ldots}}
\newcommand{\tud}[3]{#1^{\ldots #2\ldots}_{\mdots\ldots #3\ldots}}
\def \gnu {\grave{}\nu}
\def \Der {D}
\def \Derm {\Der_{-}\,}
\def \wd {\wedge}
\newcommand{\dod}[1]{\fracm{\prt}{\prt#1}}
\newcommand{\dd}[2]{\fracm{\prt#1}{\prt#2}}
\def \vep {\vc{{}\,\eps}\,}
\def \vn {\vc{{}\,n}}
\def \bara {\bar{A}}

\section{Introduction}
~~Complete description of motion of a non scalar wave in
gravitational field is given by its covariant field equations.
Under quasi-classical consideration when wave length is
sufficiently less than typical scales of observations, propagation
of the wave is substituted by motion of a particle. A.~Papapetrou
in early 50s showed that spin-gravitational interaction changes
form of the trajectory of a particle with spin. Deduction
equations of motion of the quantum particle with account of
spin-gravitational interaction can be provided by construction its
classical relativistic Lagrangian. This problem has chained a
great interest of researchers for last decades. Nevertheless a
satisfactory description was not obtained, see \cite{frszk,zr2}.

Evident progress in theoretical treatment of the problem for
photons was achieved in series of our recent articles
\cite{zr1,zr2,za,za2}. Particularly in the work \cite{za} an
approach to deduction Papapetrou equation for photon based on
field variational principle was demonstrated. Essence of that
approach was attachment the field of potential to congruence of
0-curves which specify propagation of the wave. The congruence and
the field were described by frame $\{\vn_\pm,\vn_{1,2}\}$ which is
orthonormal in sense that $<\vn_+,\vn_->=1,$
$<\vn_\alpha,\vn_\alpha>=-1,$ $\alpha=1,2$ are only non zero
scalar products of the vectors. Vector $\vn_-$ is tangent to the
0-curves everywhere and vectors $\vn_{1,2}$ constitute
polarization tangent subspace \cite{za}.  Integrand of the action
was presented as quadric form over derivatives of the potential.
Under the constraints imposed the one was reduced to form similar
to action of classic particle with an additional term describing
spin-gravitational interaction. Next it was shown that varying
this action under infinitesimal dragging changing shape of the
0-curves leads to Papapetrou equation for shape of the 0-curves.
Variation components of the potential attached to the curves
reduced equation for potential to $\Derm A^\alpha=0$. That in turn
led to equation for spin: $\Derm S_{\alpha\beta}=0,$ see
\cite{za}.

In this article we are returning to consideration of the problem
pursuing the aim to create more formal and universal approach to
deduct quasi-classical equations for motion of photon under
account of spin-gravitational interaction. At the beginning we
explain technic of variation in framework of theory of exterior
differential forms. Next we develop and improve method
demonstrated in paper \cite{za}. In particular, conception that
elements of the orthonormal frame are variables of Lagrangian
describing congruence of curves of propagation of the wave fruited
one of main ideas of the current work. That is to consider
elements of the frame as variables of gravitational field. Then to
apply infinitesimal coordinate transformations which transform
orthonormal frames, connection coefficients, potential of \elm
field  but leave invariant gravitational field. As usual Noether
theorem gives differential equations for the field observables
like stress-energy tensor (SET) and current of spin. However, this
time the equations do not exhibit covariant conservation law for
the SET, but express equation of dynamic of \elm wave.  Left hand
side (LHS) of the one is usual covariant divergence of the SET,
but in the right hand side (RHS) instead zero a term interpreted
as volume force of spin gravitational interaction arises. It
happens due to vector character of \elm field.

Then we develop our quasi-classical approach. It is based on
assumption of existance a class of locally monochromatic (LM) \elm
waves which admit geometric optical description. It means that
such the wave propagates along congruency of worldlines satisfying
an equation whose LHS coincides with equation of geodesic line in
a canonical comoving frame \cite{za}. The equation should be
derived from  the equation of dynamic. Conditions which reduce the
one to equation for shape of worldlines specify the class of LM
\elm waves. Analysis LHS of the equation of dynamic gives
conditions as follows. $T_{++}$ should be only non vanishing
component of SET of the wave and equation of continuity for
probability current had to be satisfied. In turn RHS of the
equation of dynamic should include characteristics of the particle
but not the ones of the field. It suggests that current of spin
can be expressed via tensor of spin and vector of propagation:
$S_{ab(c)}=\delta^+_c\,S_{ab},$ besides $S_{12}$ only non zero
component of the tensor of spin due to definite helicity of
photon. By other words we demand that LM \elm wave should be
transversal wave. Such the requirements are provided by
transversality of the potential and choice of suitable form of the
Lagrangian.

Analysis of form of the SET and restrictions put onto it gives
constraints which should be imposed to the potential of the field.
The constraints coincide with the ones were imposed in \cite{za}.
Condition of continuity of probability current finds its
application under deduction of equation for components of the
potential. Now $\Derm A^\alpha\neq0$ but proportional to
$A^\alpha,$ $\alpha=1,2$ and shows attenuation of the wave caused
by divergence of the beam of \elm wave. Nevertheless it does not
change our main conclusions obtained earlier \cite{za,za2}.
Specifically, under redefinition the spin as normalized $\as\nu^+$
component of current of spin equation for spin leaves the same.

\section{Lagrange formalism in
framework of orthonormal frame}

Action of \elm field in curved space-time usually expressed by
integral of Lagrangian density $\cal L$ multiplied to 4-form
$\epsilon$ of unit volume. For aims of our studies it is
convenient to express the integrand as whole 4-form $\Lambda$
which we call 4-form of Lagrangian:
\begin{equation}\label{act}
{\cal A}=\int{\cal
L}\,\epsilon=\int\Lambda,\quad\epsilon=\sqrt{-g}\,dx^0\wd\ldots\wd
dx^3=\nu^0\wd\ldots\wd\nu^3,
\end{equation}
where $\{x^i\}$ some coordinates in considered domain of
space-time and $\{\nu^a\}$ is a field of orthonormal frames dual
to $\{\vn_b\}$. In general the 4-form of Lagrangian:
\begin{equation}\label{4lag}
\Lambda=\Lambda(A_b,\Der A_c,\nu^a),
\end{equation}
depend on components of field potential $A_b(x^i)$, their
derivatives and elements of orthonormal frame $\{\nu^b\}$ which
describes gravitational field. Besides, it is convenient to
represent derivatives of the potential via covariant exterior
derivatives (CED):
\begin{equation}\label{der}
\Der A_b=(\Der_a\,A_b)\nu^a=dA_b- \con{b}{c}\,A_c,
\end{equation}
where symbol $d$ stands for exterior derivative and $\con{b}{c}$
is 1-form of Cartan' rotation coefficients. Notion of CED and
rules of operating with it should be briefly observed. We define
CED of any exterior differential form with (exterior) tensor
indexes as follows:
\def \mdots {\hspace{0.6cm}}
\begin{eqnarray*}
\Der\left(T_{\ldots a\ldots}^{\mdots\ldots b\ldots}\right)=
d\,T_{\ldots a\ldots}^{\mdots\ldots
b\ldots}+\ldots\\-\con{a}{c}\wedge T_{\ldots
c\ldots}^{\mdots\ldots b\ldots}+\ldots- \om^b_{\cdt c}\wedge
T_{\ldots a\ldots}^{\mdots\ldots c\ldots}+\ldots\;. \nonumber
\end{eqnarray*}
Usual Leibnitz rules can be easily generalized for the CED under
account of antisymmetry of exterior product of the differential
forms. We also define CED of vector:
$$\Der\vn_a=\vn_b\otimes\con{a}{b};
\quad\Der\vc\eps=\vn_a\otimes\Der_b(\eps^a)\nu^b,
$$ where symbol $\otimes$ stands for tensor product of
elements of tangent and cotangent spaces. Thanks to definitions it
is seen that rules as follows are valid:
\begin{equation}\label{rule4vt}
\vn_a\otimes\Der(T^a)=\Der(\vn_a)\wd T^a+\vn_a\otimes dT^a.
\end{equation}

In terms of 4-form of the Lagrangian the field equations have a
form as follows:
\begin{equation}\label{feq}
{\prt\Lambda\over\prt A_b}-\Der\lr\prt\Lambda\over\prt\Der
A_b\rr=0,
\end{equation}
where ${\prt\Lambda/\prt A_b}$ is a partial derivative of 4-form
of the Lagrangian over field variable $A_b(x)$. An algebraic
derivative $\prt\Lambda/\prt\Der A_b$  of 4-form $\Lambda$ over
1-form $\Der A_b$ \cite{gal} reduces it to 3-form. CED of the
algebraic derivative restores degree of the differential form up
to 4th. Notion of algebraic derivative provides compact view of
formulae especially under consideration variations of the 4-form
of Lagrangian like follows:
\begin{eqnarray}\nonumber
\Lambda(\ldots,\Der A +\delta\Der A,\ldots)- \Lambda(\ldots,\Der A
,\ldots)=\delta_{\Der A}\Lambda=\delta\Der A
\wd{\prt\Lambda\over\prt\Der A},\\
\label{set} \delta_\nu\Lambda=T_{ab}<\delta\nu^a,\nu^b>\epsilon=
T_{ab}\,{}\,\delta\nu^a\wd\!\as\nu^b=
\delta\nu^a\wd{\prt\Lambda\over\prt\nu^a};
\end{eqnarray}
where $\delta{\Der A}$ is variation of CED of potential of the
\elm field, $\delta\nu^a$ expresses variation of gravitational
field variables, $T_{ab}$ is stress-energy tensor (SET) of
electromagnetic field, asterisk stands for Hodge conjugation of
exterior differential forms.

\section{Noether theorem and Spin current}
Under infinitesimal rotations of field of the orthonormal frames
$\{\nu^b\}$:
\begin{equation}\label{lor-rot}
\delta\nu^a=-\eps^{\fdt a}_{b\cdt}\nu^b,\quad
\eps_{ab}+\eps_{ba}=0,
\end{equation}
variation of CED of the potential becomes:
\begin{equation}\label{del-der}
\delta\Der A_c=\Der\delta A_c+\delta\omega_{ab}{\prt\Der
A_c\over\prt\omega_{ab}}.
\end{equation}
Due to field equations, only part of variation of the Lagrangian
contributing to variation of the action is:
\begin{equation}\label{spin-cur}
\delta\omega_{ab}{\prt\Der A_c\over\prt\omega_{ab}}\wd
{\prt\Lambda\over\prt\Der A_c}={1\over2}\delta\omega_{ab}{\prt\Der
A_c\over\prt\omega_{[ab]}}\wd {\prt\Lambda\over\prt\Der
A_c}={1\over2}\delta\omega_{ab}\wd S^{ab}_{\cdt\cdt c}\as\nu^c,
\end{equation} where square brackets means antisymmetrization and
$S^{ab}_{\cdt\cdt c}\as\nu^c$ stands for current of spin of the
field. Variation of the connection 1-form under the considered
rotations is:
$$\delta\omega_{ab}=-\Der\eps_{ab},
$$
where
$\Der\eps_{ab}=d\eps_{ab}-\con{a}{c}\eps_{cb}-\con{b}{c}\eps_{ac}$
is CED of tensor of rotations. Substituting it into
(\ref{spin-cur}) represents variation of (\ref{act}) as follows:
$$\delta{\cal A}=\int d[\ldots]+\int\frac{1}{2}\,\eps_{ab}
\Der[S^{ab}_{\cdt\cdt c}\as\nu^c]=0.
$$ This gives us covariant equation of continuity of
3-form of current of spin:
\begin{equation}\label{eq4spin-cur}
\Der\left[S^{ab}_{\cdt\cdt c}\as\nu^c\right]=0.
\end{equation}

\section{Coordinate variation}
Infinitesimal vector field $\vc\eps $ drags coordinate
hyper-surfaces $\{x^i\}$ onto $\{y^i\}$ and elements of
orthonormal covector and vector frames $\{\vc n_a\},$ $\{\nu^b\}$
onto dragged frames $\{'\vc n_a \},$ $\{'\nu^b\}.$ Applying this
variation to action of Gravitational field gives equation of
covariant continuity for Einstein tensor in natural vector frame
and in orthonormal one:
$$G^{\fdt j}_{i\cdt|j}=\Der_b(G_{a\cdt}^{\fdt b})=0.
$$
The same procedure can be performed for (\ref{act}) as follows:
\begin{equation}\label{varOFact}\int\delta\Lambda=
\int\Lambda({}'\!\vn_0,{}'\!\vn_1,{}'\!\vn_2,{}'\!\vn_3)
{}'\!\nu^0\wd\ldots'\!\nu^3 -
\int\Lambda(\vn_0,\ldots)\nu^0\wd\ldots\nu^3,
\end{equation}
where
$\Lambda(\vn_0,\ldots)$ means value of 4-form $\Lambda$ on
4-vector $[\vn_0\vn_1\vn_2\vn_3]$. Under this definition
variations of variables of the Lagrangian are given by Lie
derivatives over vector field $\vc\eps $. Use of Cartan' first and
second structure equations expresses variations of $\nu^a$ and
$\con{a}{b}$ as follows:
\begin{eqnarray}\label{varnu}
\delta\nu^a=d\nu^a(\vep)+d\eps^a=\Der\eps^a-\con{b}{a}(\vep )\nu^b,\\
\delta\con{a}{b}=d\con{a}{b}(\vep)+d[\con{a}{b}(\vep)]
\pm[\con{c}{b}\wd\con{a}{c}](\vep)=\cur{a}{b}(\vep)
+\Der[\con{a}{b}(\vep)];\label{varom}
\end{eqnarray}
where $\cur{a}{b}=\oh R_{a\cdt cd}^{\fdt b}\,\nu^c\wd\nu^d$ is
2-form of curvature. This way variation of 4-form of Lagrangian
(\ref{4lag}) becomes:
$$\delta_A\Lambda+\delta_\nu\Lambda+\delta_{\Der A}\Lambda=
\delta A_b\dd{\Lambda}{A_b}+\delta\Der A_b\wd\dd{\Lambda}{\Der
A_b} +\delta\nu^a\wd\dd{\Lambda}{\nu^a}
$$
Noting that variation of CED as before is given by (\ref{del-der})
where this time $\delta\omega$ is given by (\ref{varom}) we after
applying (\ref{feq}) and (\ref{spin-cur}) rewrite the above
expression as follows:
\begin{equation}\label{varOF4l}
\frac{1}{2}\left\{\Om_{ab}(\vep)+\Der[\om_{ab}(\vep)]\right\} \wd
S^{ab}_{\cdt\cdt c}\as\nu^c +\Der(\eps^a)\wd
T_{ab}\as\nu^b,
\end{equation} where only first term of variation
of $\nu^a$ (\ref{varnu}) contributes owing to symmetry of the SET.
Substituting the above expression into (\ref{varOFact}) we
integrate by parts. Referring to (\ref{eq4spin-cur}) we obtain:
$$
\int d[\ldots]+\int\left\{1/2\,\Om_{ab}(\vep)\wd S^{ab}_{\cdt\cdt
c}\as\nu^c-\eps^a\Der\left[T_{ab}\as\nu^b\right]\right\}=0.
$$
It brings us equation as follows:
\begin{equation}\label{vf}
\vn_c\otimes\Der\left[T^c_{\cdt b}\as\nu^b\right]
=\oh\vn_c\otimes\,R_{ab\cdt\cdt}^{\fdt\fdt cd}\,S^{ab}_{\cdt\cdt
d}\,\epsilon.
\end{equation}
The one in contrast with analogous equations for scalar fields is
not equation of covariant conservation of the SET due to non zero
RHS appeared. The RHS expresses volume force of spin-gravitational
interaction while the LHS exhibits CED of current of
stress-energy-momentum. Because of this we will call (\ref{vf}) as
equation of dynamic of the \elm wave in curved space-time.
Accordingly to rule (\ref{rule4vt}) LHS of (\ref{vf}) can be
represented as:
\begin{equation}
\vn_c\otimes D\left[T^c_{\cdt b}\as\nu^b\right]=D(\vn_c)\wd
T^c_{\cdt b}\as\nu^b+\vn_c\otimes d\left[T^c_{\cdt b}\as\nu^b
\right].\label{cdset}
\end{equation}
It should be marked that use of the above mentioned rule becomes
possible owing to fact that all indexes in the above expression
are dummy. Such form of presentation of covariant divergence of
the SET is convenient for transition to quasi classical
consideration.

\section{Quasi classical approach, Schweber Lagrangian and constraints}
\label{sec}

Conception of our quasi classical description is a deduction of
equation for vector of propagation $\vn_-$ from the equation of
dynamic (\ref{vf}). Besides we demand that obtained equation
becomes coincide with equation of geodesic line under switch off
the spin. This means that covariant divergence of SET should be
reduced to covariant derivative of $\vn_-$. Equality (\ref{cdset})
shows that this result is provided by constraints as follows:
\begin{eqnarray}\label{tppdom}
T_{++}\,{}\,\mbox{\it is only non vanishing component of the SET,}\\
\hspace{2.75cm}d\left[T_{++}\,\as\nu^+\right]=0.
\hspace{2.75cm}\label{prob}
\end{eqnarray}
It should be distinguished that (\ref{tppdom},\ref{prob})
characterize subspace of locally monochromatic (LM) \elm waves and
class of canonical comoving frames \cite{za} which allows most
convenient form of quasi-classical equations for the waves.  The
same form of SET has monochromatic \elm wave in flat space-time
propagating along $\vn_-$:
$$T^{ab}=\delta^a_-\,\delta^b_-\,\omega^2|\va|^2,
$$ where $\omega$ is frequency of the wave.
Thus (\ref{prob}) expresses continuity of current of probability
in case of flat space-time. Under this conditions LHS of
(\ref{vf}) becomes:
$$\Der\left[\vn_-\right]\wd T^-_{\csdt+}\as\nu^+=
T^-_{\csdt+}\Der_-\left[\vn_-\right]\,\epsilon.
$$
This way we reduce (\ref{vf}) to sought form:
\begin{equation}\label{dyneq}
T_{++}\Der_-\left[\vn_-\right]=\vn_c\,R^{\fdt c}_{d\cdt
ab}\,S^{abd}.
\end{equation}
LHS of the equation coincides with geodesic equation for vector
field $\vn_-,$ however in the RHS stands a term including
contraction of the curvature with current of spin of the field
which we call force of spin-gravitational interaction. Besides the
force still is written in terms of field theory. To pass to
complete quasi classical description we need to determine field
Lagrangian.

Under study process of propagation of LM \elm wave it is relevant
admit Schweber Lagrangian \cite{Mic} which provides adequate
quasi-classical description of LM wave:
\begin{equation}\label{Schweber}{\cal L}=1/2\,\Der_a A^c\,\Der_b A_c
<\nu^a,\nu^b>.
\end{equation}
In terms of orthonormal frames we write 4-form of the Lagrangian
as follows:
\begin{equation}\label{L4} \Lambda=\oh\Der
A_c(\vn_a)\left\{\Der A^c\wd\as\nu^a\right\}.
\end{equation}
Now let's calculate variation of the action under variation of
elements of the frame:
\begin{equation}\label{dnuL}\delta_\nu\Lambda=\oh\Der
A_c(\delta\vn_a)\,\Der A^c\wd\as\nu^a+ \oh\Der A_c(\vn_a)\,\Der
A^c\wd\delta\as\nu^a.
\end{equation}
It is evident that: $\delta\vn_a=-\delta\nu^b(\vn_a)\vn_b$. So
first term in (\ref{dnuL}) can be rewritten as follows:
$$-\oh<\delta\nu^b,\nu^d>\Der_bA_c\,\Der_dA^c.
$$ During variation of the second term asterisk conjugation should be
varied:
\begin{eqnarray*}\delta\as\nu^a=1/3!\,\eps^a_{\cdt
bcd}\,\delta(\nu^b\wd\nu^c\wd\nu^d)= \oh\eps^a_{\cdt
bcd}\,\delta\nu^b\wd\nu^c\wd\nu^d\\\vphantom{\int^\int}
\Rightarrow\nu^e\wd\delta\as\nu^a=-\oh\eps^a_{\cdt
bcd}\eps^{ecd}_{\cdt\cdt\cdt f}\,\delta\nu^b\wd\as\nu^f=
(\eta^{ae}\eta_{bf}-\delta^a_f\delta^e_b)<\delta\nu^b,\nu^f>\epsilon.
\end{eqnarray*} This brings us:
$$\delta_\nu\Lambda=-\left(\Der_aA_c\,\Der_bA^c-
\oh\Der_eA_f\Der^eA^f\,\eta_{ab}\right)<\delta\nu^a,\nu^b>\epsilon
$$ By other words, accordingly to (\ref{set}), we find:
\beql{theset} T_{ab}=-\left(\Der_aA_c\,\Der_bA^c-
\oh\Der_eA_f\Der^eA^f\,\eta_{ab}\right).\eeq

The next part of the variation contributing to RHS of (\ref{vf})
is variation over Cartan' rotation 1-form. The one appears in
expression of CED of potential (\ref{der}). Evidently
$\delta_\omega\Der A_c=$ $-\delta\con{c}{d}A_d$, so
$$\delta_\omega\Lambda=\delta_\omega\Der A_c\wd\dd{\Lambda}{\Der
A_c}=\delta\con{c}{d}\wd A_d\Der_b
A^c\as\nu^b=\oh\delta\omega^{cd}\wd A_\lcom{d}\Der_b
A_\rcom{c}\as\nu^b.
$$ Now equating this expression to (\ref{spin-cur})
we obtain:
\beql{spcur}S_{cdb}=A_\lcom{d}\Der_b A_\rcom{c}.\eeq

Since photons has definite helicity they are presented by \elm
waves with circular polarization \cite{za}. Hence it is convenient
to consider complex valued amplitudes of the potential which
describe phase shift of circularly polarized waves. For this aim
an ordinary procedure of redefinition the observables being
quadric forms of the potential components is used. That is
$A_pA_q\to\oh\bar A_\lacm{p}A_\racm{q},$ where curly brackets mean
symmetrization over the indexes. After applying it to expressions
for the SET and the current of spin we obtain their form in
complex amplitudes:
\begin{eqnarray}
T_{ab}=-{1\over2}\left\{\Der_a\bar A_c\,\Der_b
A^c+\Der_aA_c\,\Der_b\bar A^c-\eta_{ab}\,{}\,\Der_c\bar
A_d\,\Der^cA^d\right\},
\label{setcom}\\
S_{cdb}={1\over2}\left(\bar A_\lcom{d}\Der_b A_\rcom{c}+
A_\lcom{d}\Der_b\bar A_\rcom{c}\right). \label{spccom}
\end{eqnarray}

Now let us find constraints should be put to the potential. As we
consider LM \elm wave we expect that potential of the wave has a
structure as follows:
\begin{eqnarray}\label{potstruc}
\va=\vc a\,e^{i\phi},\quad
\Der_+\phi=\omega,\quad\Derm\phi=0,\quad\Der_+\vc a=0\\
\Rightarrow\label{derpA}\Der_+A_\alpha=i\omega A_\alpha.
\end{eqnarray}
Due to (\ref{setcom}) it is seen that condition of dominance of
$T_{++}$ demands vanishing derivatives of the potential along
polarization vectors:
\begin{equation}\label{derpol}\Der_\alpha
\va=0,\,{}\,\alpha=1,2.
\end{equation}
Vanishing the derivative along direction of propagation $\vn_-$
makes the condition surely satisfied, however it is not only
possibility as it will be shown later. Anyhow value of $T_{++}$
can be calculated now due to fact that $\eta_{++}=0$. Under
constraints imposed it is reduced to the form of $T_{++}$ of
monochromatic wave in flat space-time:
\begin{equation}T_{++}=-\omega^2<\bar{\va} ,\va>=\omega^2|\va|^2
\label{Tpp}
\end{equation} Quasi classical picture assumes that spin of photon
moves along field of vectors $\vn_-$ and has only $S_{12}$
component due to fact of definite helicity of photon. Constraints
(\ref{derpol},\ref{derpA}) provide validity of the first condition
which can be written as $S_{cdb}=S_{cd}\,\eta_{b-},$ where
$S_{cd}$ is tensor of spin. The second condition is supplied by
additional constraints:
\begin{equation}\label{apm}
A_{\pm}=0.
\end{equation}

\section{Field equations}
Under varying Lagrangian (\ref{Schweber}) over components of
potential we obtain reduced form of Maxwell equations:
$$\Der\left[\Der_bA^c\,\as\nu^b\right]=0.
$$
Let's consider equations for nonzero components of the potential.
After imposing constraints we obtain:
\beql{derpm}\Der_\lacm{+}\left(\Der_\racm{-}A^\alpha\right)
\epsilon+\sum_{c=\pm}\Der_cA^\alpha\,d\as\nu^c=0.\eeq As usual
under quasi classical (geometric optical) consideration we assume
that values of second derivatives of amplitudes of the potential
should be neglected:
$$\omega^{-1}|\Der_a(\Der_b\, a^c)|
=\omega^{-1}O\left(|\vc a |^{-3}\left[\Derm|\vc a
|^2\right]^2\right)=o\left(|\vc a |^{-1}\,\Derm|\vc a|^2\right).
$$ It reduces (\ref{derpm}) to:
\begin{equation}\label{ini4fe}
\left[2i\omega\Derm+i(\Derm\omega)\right]A^\alpha\,\epsilon+
i\omega\,A^\alpha\,d\as\nu^++(\Derm A^\alpha)\,d\as\nu^-=0.
\end{equation}
Complex conjugation of (\ref{ini4fe}) gives equation for complex
amplitude. Let's consider contraction of (\ref{ini4fe}) with $\bar
A_\alpha$ and contraction of the one complex conjugated with
$A_\alpha$. Difference between them gives:
\begin{eqnarray*}
i\left[\omega\Derm<\bar{\va}
,\va>+(\Derm\omega)<\bar{\va},\va>\right]\epsilon
+i\omega<\bar{\va},\va>\,d\as\nu^++\\
+\oh\left[\bara_\alpha\Derm
A^\alpha-A_\alpha\Derm\bara^\alpha\right]\,d\as\nu^-=0.
\end{eqnarray*}
Due to structure of potential and constraints we daresay that last
term in the above equation is zero. This manner we rewrite the
equation as follows:
$$\omega^{-1}d[\omega^2<\bar{\va},\va>\as\nu^+]
-(\Derm\omega)<\bar{\va},\va>=0.
$$
But owing to (\ref{Tpp}) and (\ref{prob}) first term in the
equation is zero. It gives $\Derm\omega=0$. By other words value
of derivative of the frequency in chosen frame along vector
$\vn_-$ is vanishing in quasi classical approximation:
\begin{equation}\label{derom}
\Derm\omega=O\left(\left[\Derm<\bar{\vc a },\vc a>\over <\bar{\vc
a },\vc a>\right]^2\right)=o\left(\omega\,\Derm<\vc a ,\bar{\vc a
}>\over<\vc a ,\bar{\vc a }>\right).
\end{equation}
This result allows to rewrite (\ref{ini4fe}) and (\ref{prob}) as
follows:
\begin{eqnarray}\label{redEQ4pot} i\omega\left[2\Derm
A^\alpha\epsilon+A^\alpha\,d\as\nu^+\right] +(\Derm A^\alpha)\as
d\as\nu^-=0,\\\label{RedProb}
d\left[<\va,\bar{\va}>\as\nu^+\right]=0.
\end{eqnarray}
Equating to zero factor at $\omega$ in (\ref{redEQ4pot}) as it is
used to do under geometric optical approximation we obtain
equation for potential components. In turn (\ref{RedProb})
expresses conservation of probability current of the wave and
allows to exclude $d\as\nu^+$ from the equation for the potential:
\begin{equation}\label{eq4pot}
\Derm A^\alpha=\oh A^\alpha\,\as
d\as\nu^+=A^\alpha{\Derm(\bar{A_c}A^c)
\over2\bar{A_c}A^c\vphantom{{A^2}^2}}.
\end{equation}
Equation (\ref{eq4pot}) describes attenuation of the amplitude of
potential caused by divergence of worldlines of propagation of the
wave.

\section{Explicit form of SET, current of spin and Papapetrou equation}
\label{s7}

Substituting  (\ref{eq4pot}) together with constraints into
(\ref{setcom},\ref{spccom}) allows us to obtain an explicit form
of elements of SET:
\begin{eqnarray*}
\Der_a\bar{A_b}\,\Der^aA^b=\Der_\lacm{+}\bar{A_b}\,\Der_\racm{-}A^b=
[-i\omega\,\bar{A_b}\,A^b/2+1/2\bar{A_b}\,i\omega A^b]
\fracm{\Derm\bar{A_c}A^c}{\bar{A_c}A^c}=0,\\
T_{+-}=-\oh\Der_\lacm{+}\bar{A_c}\,\Der_\racm{-}A^c=0,\quad
T_{--}=-\Derm\bar{A_b}\,\Derm A^b\ll T_{++}.
\end{eqnarray*}
The calculations confirm assumptions about view of elements of SET
of LM wave. Next step is to calculate elements of spin tensor.
Thanks to constraints all its elements with third index different
from "$\pm$" is surely zero. But accordingly to (\ref{spccom}):
\begin{equation}\label{ffsc}
\begin{array}{rcl}
S_{cd+}&=&i\omega\,\bara_\lcom{d}\,A_\rcom{c}=:S_{cd},\\
S_{cd-}&=&\frac{\Derm|\va|^2}{|\va|^2}
\left(\bara_\lcom{d}A_\rcom{c}+A_\lcom{d}\bara_\rcom{c}\right)=0,
\vspace{0.15cm}\\
\Rightarrow\quad S_{cdb}&=&\delta^+_b\,S_{cd}.\\
\end{array}
\end{equation}
It approves expected factorized form of current of spin. Now
substituting (\ref{Tpp}) and (\ref{ffsc}) into (\ref{dyneq}) we
obtain equation for vector $\vn_-$ being tangent to worldlines of
propagation of the wave:
\begin{equation}\label{papa}
\omega^2|\va\,|^2\,\Derm\vn_-=\oh\vn_c\,R^{abc+}\,S_{ab}.
\end{equation} In fact the one coincides with Papapetrou
equation for trajectory of photon obtained us in work \cite{za},
although combinatorial factor $\oh$ was not present there. It
happened only by a variety in the definitions of tensor $S_{ab}$
in the articles. Under practical calculations it is more
convenient to introduce normalized tensor of spin \cite{za2}:
\begin{equation}\label{nsp}
\sigma_{ab}=\frac{S_{ab}}{\omega|\va\,|^2}=
\frac{\bara_\lcom{a}A_\rcom{b}}{i|\va\,|^2}.
\end{equation}
As expected, the one is transporting parallel itself in
polarization tangent subspace:
\begin{eqnarray*}
\Derm\sigma_{\alpha\beta}=i\left[
\fracm{\Derm(\bara_\lcom{\alpha})A_\rcom{\beta}+
\bara_\lcom{\alpha}\Derm(A_\rcom{\beta})}{\bara_cA^c}-
\fracm{\bara_\lcom{\alpha}A_\rcom{\beta}\,\Derm(\bara_cA^c)}{(\bara_dA^d)^2}
\right]=\\
=i\left[
\fracm{\Derm(\bara_cA^c)}{2(\bara_dA^d)^2}\,\{\bara_\lcom{\alpha}A_\rcom{\beta}+
\bara_\lcom{\alpha}A_\rcom{\beta}\}-
\fracm{\bara_\lcom{\alpha}A_\rcom{\beta}\,\Derm(\bara_cA^c)}{(\bara_dA^d)^2}
\right]=0.
\end{eqnarray*}
Finally excluding potentials of the \elm field, we write system of
Papapetrou equations for vector of propagation of LM
electromagnetic wave and its spin as follows:
\begin{equation}\label{Papapetrou}\Derm\sigma^{\alpha\beta}=0,
\hspace{1.5cm}\omega\,\Derm(\vn_-)= \oh\vn_c\,R^{\fdt\fdt\,c
}_{\alpha\beta\cdt-}\, \sigma^{\alpha\beta};
\end{equation}
where value of frequency $\omega$ leaves constant in canonical
comoving frame along the trajectory with accuracy up to second
order in accordance with \cite{za}.

\section{Remarks}


It should be underlined that idea of this work is grounded on
conception of article \cite{za}. Specifically, in the above cited
paper we started by consideration gauge invariant Lagrangian
${\cal L}=\oh\Der_aA_b\,\Der_cA_d\,<\nu^a\wd\nu^b,\nu^c\wd\nu^d>$.
Next we expand scalar products as
$<\nu^a,\nu^\lcom{c}><\nu^b,\nu^\rcom{d}>$ and put the
constraints. It was guessed that only terms $<\!\nu^a,\nu^c\!>
<\!\nu^b,\nu^d\!>$ would contribute to variation of the action due
to fact that they yields physically interpretable expressions. In
present work that assumption finds its expression in choice of
form of Lagrangian. From another point of view selection of
Schweber Lagrangian together with the constraints serves
transversality of LM \elm wave.

We also note that variation of omitted scalar product in Schweber
Lagrangian
$\Der_aA_c\Der_bA^c\to\Der_aA_c\Der_bA_d\,<\nu^c,\nu^d>$ may
contribute only to the SET. Calculations with account of the
constraints and symmetrization over complex amplitudes analogous
to the ones performed in section \ref{s7} shows that the
contribution vanishes.

Equation for probability current (\ref{RedProb}) can be rewritten
as:
$$\Derm<\va,\bar{\va}>+<\va,\bar{\va}>\,Div\,\vn_-=0,$$
where $Div$ stands for covariant divergence of the vector. Under
our approximation we daresay that both terms in the equation are
small but not vanishing. But in our former work \cite{za} it was
implicitly assumed that (\ref{RedProb}) is satisfied by vanishing
of both of its terms. Hence equation for amplitudes of the
potential obtained us earlier is particular case of
(\ref{eq4pot}). The last is realized when worldlines of
propagation of the wave are parallel themselves locally, as it was
demanded in the previous article.

\section*{Acknowledgment}

Author thanks to Professor Z. Ya. Turakulov who involved the
author to the studies and provided the work by his help.

\end{document}